%% file: main.tex
\documentclass[aps,llncs,preprint]{revtex4-1} 
\usepackage{graphicx}
\usepackage{amsmath}
\usepackage{float}
\usepackage[colorlinks,allcolors=blue]{hyperref}
\usepackage{color}
\usepackage{soul}
\usepackage{subfiles}

\begin{document}

\subfile{myfirstfile}

\pagebreak
\subfile{mysecondfile}

\end{document}

%% file: myfirstfile.tex
\title {Spectroscopic signature of anisotropic order parameter in Kagome lattice superconductor LaRh$_3$B$_2$}

\author{Mona Garg}
\email{mona@iisermohali.ac.in}
\author{Nikhlesh S. Mehta}
\author{ Ghulam Mohmad}
\author{Savita Chaudhary}
\author{Yogesh Singh}
\author{Goutam Sheet}

\email{goutam@iisermohali.ac.in}
\affiliation{Department of Physical Sciences, Indian Institute of Science Education and Research Mohali, S.A.S. Nagar, Punjab, India}

\begin{abstract}
The physics of the Kagome metal LaRh$_3$B$_2$ along with its superconductivity below 2.6 K, unlike other popular Kagome metals, is not known to be significantly influenced by the electron correlations. While the indirect techniques to probe the bulk superconducting properties of  LaRh$_3$B$_2$ indicate a conventional isotropic order parameter, we show that the direct spectroscopic determination of the superconducting energy gap reveals an anomalous suppression of Andreev reflection between LaRh$_3$B$_2$ and a normal metal. This observation hints to the presence of incomplete superconducting gap formation, at least along certain momentum directions, and consequent low-lying quasiparticle states. An analysis of multiple Andreev reflection spectra captured at different points on the surface of LaRh$_3$B$_2$ reveals a distribution of the superconducting energy gap which is consistent with an anisotropic superconducting order parameter.

\end{abstract}
\maketitle

\section{Introduction}
A central theme of condensed matter physics is to search for novel quantum phases of matter in solid state systems. A Kagome lattice is composed of hexagons and triangles in a network of corner-shared triangles and such a lattice has the potential to host physical effects such as geometric frustration, nontrivial band topology, van Hove singularities (vHSs), flat bands etc. This makes the Kagome systems a fertile platform to study the interplay of lattice, topology, magnetism and electron correlation effects \cite{liu_magnetic_2019, kang_dirac_2020, liu_orbital-selective_2020, cable_magnetic_1993, nakatsuji_large_2015, hou_observation_2018, ma_rare_2021, pasco_tunable_2019, zheng_quantum_2024, zhang_electronic_2022, guguchia_hidden_2023, guin_zerofield_2019, zhang_topological_2017, liu_giant_2018, lee_quantum-spin-liquid_2007}. A variety of phases have already been discovered in various Kagome materials, such as Mott insulator, quantum spin liquid, topological semimetal and insulator and more recently charge density wave orders. In this context, more recently, several Kagome metals have gained attention as many of these materials also undergo a superconducting phase transition at low temperatures. Due to the presence of flat bands in these Kagome metals, electrons in these bands have a large effective mass and a very small band velocity. This indicates that many-body effects along with strong correlation effects are some of the important characteristics of these systems. In addition, the presence of flat bands near the Fermi energy may potentially result in high superconducting transition temperatures as well as strong superconducting fluctuations (SCFs). Therefore, many theoretical and experimental studies have been initiated on the relationship between flat bands and superconductivity to unravel the exotic physical \& electronic properties of the Kagome superconductors. For example, the topological Kagome metals AV$_3$Sb$_5$ (A = K, Rb, Cs) family of materials show superconductivity, charge density waves (CDW), nematic/stripe order,pair-density wave phase (PDW), large anomalous Hall effect (AHE) and multiple van Hove singularities (VHS) near the Fermi energy ($E_F$), all in a single material, as a function of temperature \cite{zhao_cascade_2021, chen_roton_2021, ortiz_cs_2020, yu_concurrence_2021, wen_emergent_2023, wu_nature_2021, kang_twofold_2022}. 

Here we focus on another family of superconductors RT$_3$X$_2$ (R = lanthanide, T = 4\textit{d} or 5\textit{d} transition metal, X = Si, B) that are described by the Kagome lattice structure. These materials are proposed to host Dirac cone, Flat bands as well as VHSs \cite{okubo_unique_2003}. There are several reports of superconductivity and ferromagnetism in such compounds \cite{gui_lair_2022, gong_superconductivity_2022, wang_quantum_2023}. Several unconventional properties have been reported in LaRu$_3$Si$_2$ and CeRu$_3$Si$_2$, possibly arising from electron correlations from the flat bands. Therefore, it is most important to investigate the superconducting phase of this family of Kagome superconductors by spectroscopic tools. \cite{rauchschwalbe_superconductivity_1984, mielke_nodeless_2021, li_anomalous_2011}. In this paper, we present point contact Andreev reflection (PCAR) spectroscopy experiments of LaRh$_3$B$_2$ to study the nature of its superconducting order parameter. Our PCAR experiments on the material gives a direct spectroscopic measurement of the superconducting gap and also reveals an anomalous suppression of Andreev reflection on LaRh$_3$B$_2$. From a detailed analysis of the PCAR data we show that the superconducting phase in LaRh$_3$B$_2$ has an anisotropic character.

\section{Experiments} 
The low-temperature measurements were performed on polycrystalline LaRh$_3$B$_2$ where multiple single crystallites with randomly oriented facets coexist on the surface. The materials were synthesized by arc-melting stoichiometric ratios of La (3N, Alfa Aesar), Rh (5N, Alfa Aesar), and B (6N, Alfa Aesar). The melted buttons were flipped over and melted 5–10 times more to promote homogeneity. Powder x-ray diffraction confirmed that the synthesized samples are in single-phase. The lattice parameters estimated after refinement are in  great agreement with that available in the published literature. The details of materials growth and characterizations are reported elsewhere\cite{chaudhary_role_2023}.

The Andreev reflection spectroscopic measurements were performed by measuring the transport characteristics of ballistic point contacts between superconducting LaRh$_3$B$_2$ and normal metallic Ag tips. The measurements were performed by using a home-built point-contact spectroscopy(PCS) probe in a He3 cryostat with a sample temperature of around 480 mK\cite{10.1063/1.5119372}. First, the surface of LaRh$_3$B$_2$ was polished and mounted on 0.1 mm thick cover glass using G-varnish (CMR direct). Then it was mounted on copper disc which is used as the sample holder in the home-built PCS probe using G-varnish for heat-sinking. A calibrated Ruthenium Oxide (RuOx) thermometer and a resistive heater were attached to the same holder for accurate measurement and control of temperature. The sample was placed at the center of a $z$-axis solenoidal superconducting magnet with magnetic field upto 7 Tesla. The mesoscopic point contacts were made by gently engaging a sharp Ag tip using a piezo based nano-positioner on the surface of LaRh$_3$B$_2$. A conventional Lock-in based ac modulation technique by sweeping a dc current superimposed on the small amplitude ac current operating at 671 Hz was used to measure differential conductance ($dI/dV$) as a function of applied dc voltage ($V$) in a four probe geometry.

\section{Results and discussion}
In the electrical transport through a ballistic point contact between two conductors the resistance of the contact depends on contact diameter and is independent of global disorder. In such a contact, the contact diameter (\textit{d}) is less than the elastic mean free path (\textit{l}) of electrons in either of the conductors. The voltage $V$ applied across such contacts defines the energy of the electrons within the contact region. Since ballistic transport doesn't involve conventional scattering, the electron that accelerates under the applied $V$ can attain very high energy and can resonantly excite the elementary excitation modes in the materials forming the contact. This leads to non-linearities in the $I$-$V$ characteristics\cite{jansen_point-contact_1980}. Further, the ballistic transport characteristics across normal metal-superconductor junctions are dominated by Andreev reflection, a quantum process. Ideally, Andreev reflection should cause an enhancement of the differential conductance $dI/dV$ when the electron energy is less than the superconducting energy gap ($\Delta$) by a factor of 2. In real experiments, this factor is measured to be slightly less than 2 due to multiple effects such as a non-zero temperature, finite interfacial barrier\cite{PhysRevB.25.4515}, two-level fluctuations\cite{10.1063/1.3479927}, and finite quasiparticle lifetime\cite{PhysRevB.49.10016} etc. A $dI/dV$ vs energy ($E$) spectrum thus obtained across a ballistic point contact between a normal metal and a superconductor is traditionally analyzed by a generalized theoretical model by Blonder-Tinkham- Klapwijk (BTK)\cite{PhysRevB.25.4515} that assumes the interface as a delta potential barrier ($V_0\delta(x)$) with the barrier strength described by a dimensionless parameter $Z = V_0/\hbar v_F$ where $v_F$ is the Fermi velocity of the material under investigation. When $Z$ is small ($\sim 0.5$), two peaks symmetric about $V =0$ appear. Such peaks are the hallmark signatures of Andreev reflection. This highlights the significance of PCAR spectroscopy as a powerful tool for studying all type of (conventional and unconventional) superconductors.
\begin{figure}[h!]
    \includegraphics[width=.9\textwidth]{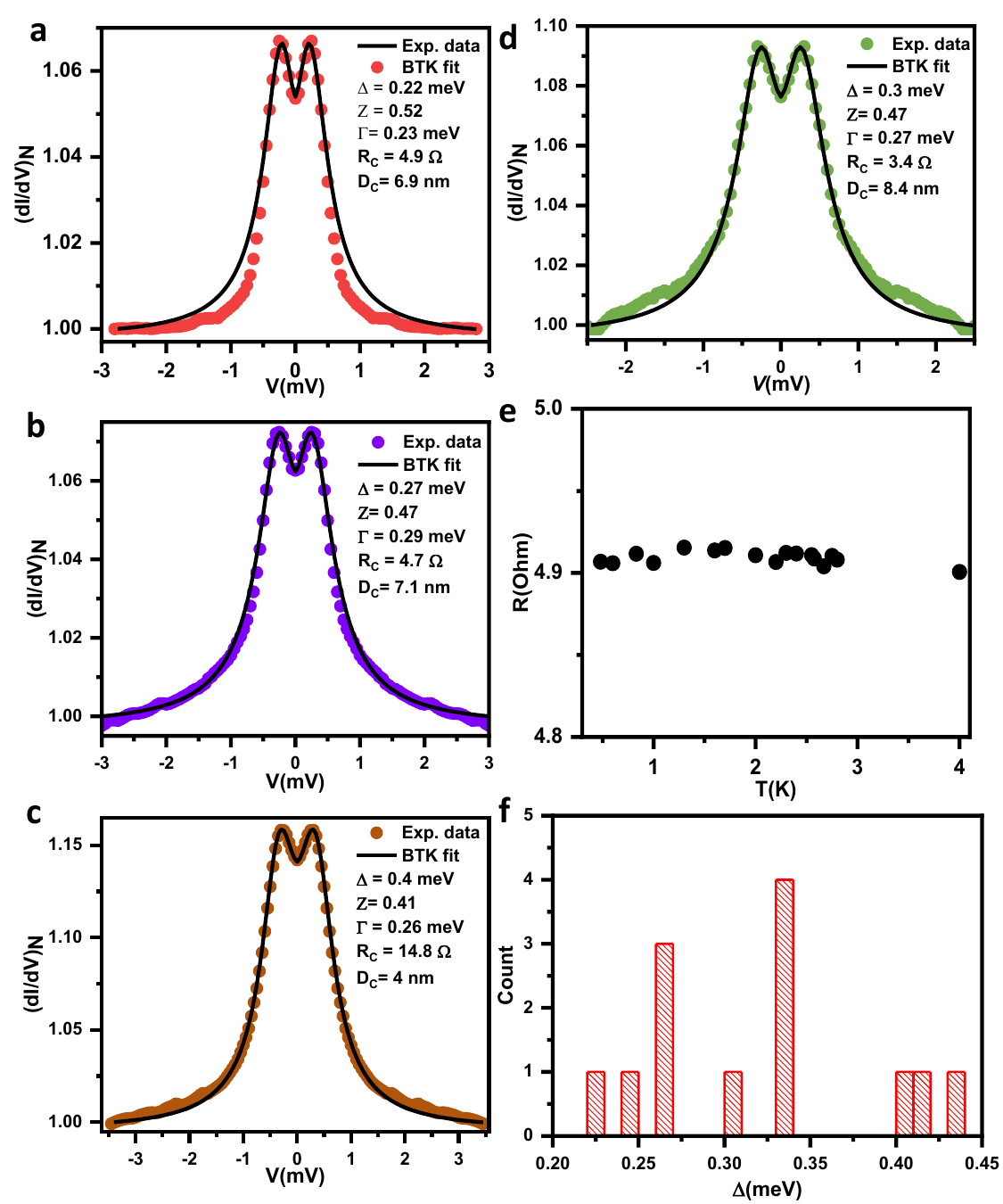}
    \caption{(a-d) Conductance spectra obtained in the ballistic regime (solid circles) and corresponding single gap BTK fit (black line). All the spectra were recorded at 0.48 K in the absence of a magnetic field. The extracted fitting parameters $\Delta$, $\Gamma$ and $Z$ are also mentioned for each spectrum. (e) Temperature dependence of normal state resistance (R$_N$) of the spectrum shown in (a). (f) Statistics of the fitting parameter $\Delta$ for 13 independent point-contact spectra. }
\end{figure}

\begin{figure}[h!]
    \centering
    \includegraphics[width=.75\textwidth]{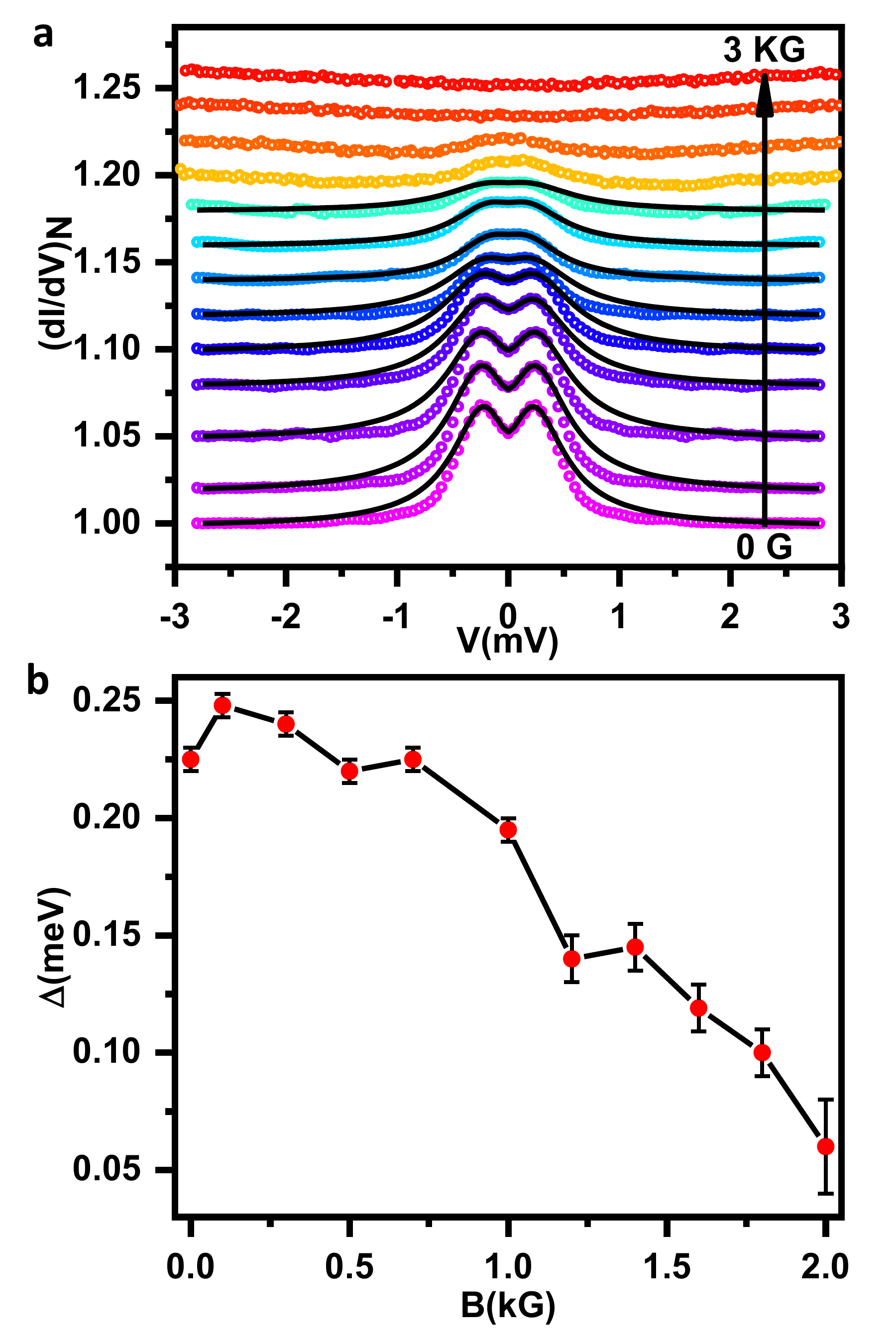}
    \caption{(a) Magnetic field ($H$) dependence of the conductance spectra (colored circles) with BTK fit (black lines) at 0.48 K. (b) Variation of superconducting gap ($\Delta$) with magnetic field.}
\end{figure}
Figures 1(a-d) show four representative PCAR spectra recorded from point contacts at four different regions on the surface of LaRh$_3$B$_2$. The spectra are first normalized to the conductance at the high bias (normal state conductance). Here, the double peaks symmetric about $V =0$, the hallmark signature of Andreev reflection, are clearly seen. The solid black lines in Figures 1(a-d) represent the theoretically simulated spectra as per the modified BTK theory with the barrier strength $Z$, the broadening parameter $\Gamma$ and the superconducting gap $\Delta$ as the fitting parameters. It is clearly seen that while some spectra are well-fitted over the full energy range, others show a deviation from the BTK theory beyond a range over $\pm$ 0.7 mV. Figure 1(f) shows a distribution of $\Delta$ obtained from 14 spectra obtained at different points on LaRh$_3$B$_2$ surface. Please refer to Figures 2 and 3 in the supporting information file for the data along with the BTK fits. The median value of $\Delta$ is 0.33 meV. The estimated $\Delta$(0)/k$_B$T$_c$ from this analysis is $\sim$ 1.43 which is smaller than the weak-coupling BCS value.

As mentioned above, one of the key observations is that several spectra captured at different points on the surface of LaRh$_3$B$_2$ were not described well within the BTK framework over the entire energy range. While the low-bias portions of those spectra matched well with the BTK theory, the spectra deviated from the BTK theory at the higher bias. In addition, the fact that the amplitude of the superconducting energy gap revealed by these spectra is significantly smaller than the gap estimated by indirect bulk measurements ($\sim$ 0.52 meV)\cite{chaudhary_role_2023}. Before analyzing such features, as a first check, it is important to confirm that the observed deviation is not due to thermal effects. We noted that the spectra display Andreev reflection driven double peak structures prominently and no extra (anomalous) features like conductance dips\cite{PhysRevB.69.134507, kumar_nonballistic_2021} are observed. This suggests that the point contacts are in the ballistic (or diffusive) regimes of transport where true spectroscopic parameters can be obtained. Furthermore, the normal state resistance of the point contacts varied from 1.5 $\Omega$ to 15 $\Omega$ and the contact diameters (calculated using Wexler’s formula) varied between 4 nm $\sim$ 12.7 nm. This is considerably smaller than the mean free path $\sim$ 43 nm at 4 K. Also, as a representative illustration, the variation of normal state resistance of spectrum 1(a) with temperature is shown in Figure 1(e) where it can be seen that the normal state resistance of the spectra remained temperature independent. All these observations collectively confirm that all the spectra presented in this paper are in the ballistic regime of transport and consequently, the heating effects are not considerable.

\begin{figure}[h!]
    \centering
    \includegraphics[width=.75\textwidth]{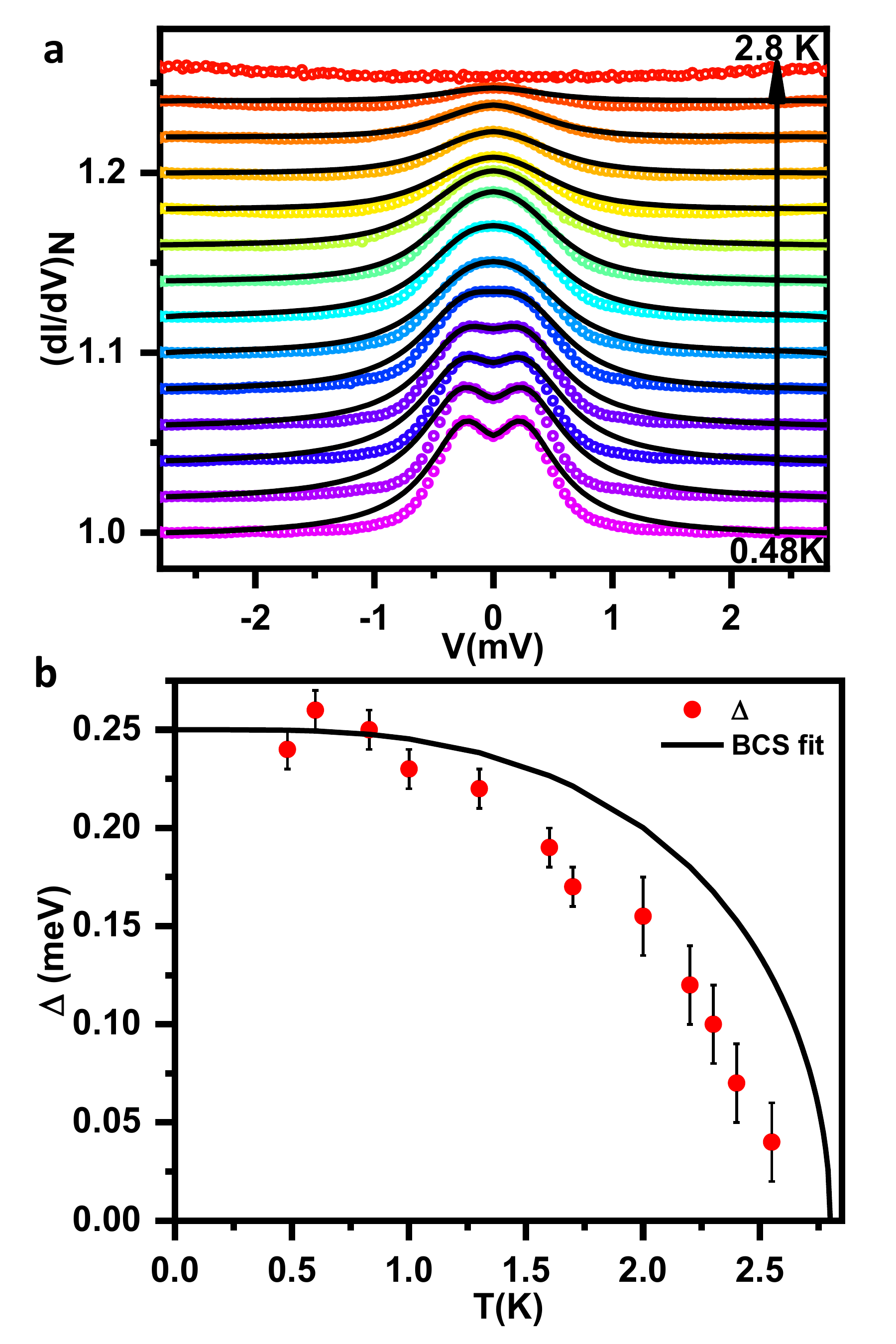}
    \caption{(a) Temperature($T$) dependence of the conductance spectra (colored circles) with BTK fit (black lines) at 0 kG. (b) Variation of superconducting gap ($\Delta$) with temperature.}	
\end{figure} 

The other important fact that is immediately discernible in the spectra is that the conductance enhancement due to Andreev reflection is dramatically suppressed. While under ordinary circumstances, involving conventional superconductors, the zero-bias conductance should undergo a nearly 200\% enhancement upon Andreev reflection, the spectra on LaRh$_3$B$_2$ display an enhancement that varied between 6\% and 25\%. In order to gain an overall understanding of the anomaly discussed above, we investigate the evolution of the PCAR spectra with magnetic field and temperature. For the magnetic field dependent experiments, a magnetic field was applied along the $z$-axis (vertical). The $dI/dV$ spectra showed a gradual disappearance of the features associated with superconductivity with increasing magnetic field (Figure 2(a)). In the modified BTK theory with additional broadening parameter/quasiparticle lifetime $\Gamma$, the $dI/dV$ spectra were well described. The BTK simulated spectra are shown as black lines in Figure 2(a). The evolution of $\Delta$ with magnetic field is shown in Figure 2(b), where the energy gap is seen to decrease monotonically with applied field strength. The overall spectroscopic features corresponding to Andreev reflection for these point contacts ceased to exist beyond a critical field of $\sim$ 3 kG. A representative temperature dependence is shown in Figure 3(a). As expected for superconductors, the spectral features gradually diminished with increasing temperature and the overall spectroscopic features disappeared around $T\sim$ 2.8 K. To note this temperature is the same as the bulk $T_c$ of LaRh$_3$B$_2$. This further supports the claim that contact heating in the presented measurements is not significant. In Figure 3(b), we show the plot of $\Delta$ (red dots) with $T$ where $\Delta$ was obtained by comparing the temperature dependent experimental spectra with the BTK simulations (shown as black lines on the temperature dependent experimental curves in Figure 3(a)). We also show the expected temperature dependence of $\Delta$ (the solid black line in Figure 3(b)) within the BCS theory of conventional superconductivity. It is seen that there is a deviation of the experimentally measured $\Delta$ from the BCS line. This deviation further indicates that the order parameter in LaRh$_3$B$_2$ might have an anisotropic character. To establish whether the anisotropy arises from an an anisotropic density of states or an anisotropy in the electron-phonon coupling is beyond the scope of the PCAR spectroscopy. There is also a possibility of an anisotropic order parameter caused by multiple gaps forming on distinct Fermi sheets. However, our PCAR experiments did not resolve a second gap in the system.

Before discussing the details of the possible type of unconventionality, it is important to first take note of the unique Fermi surface properties of LaRh$_3$B$_2$. Earlier, high-resolution angle-resolved photoemission spectroscopy (ARPES) and de Haas–van Alphen (dHvA) experiments confirmed that LaRh$_3$B$_2$ is a compensated metal with both electron and hole Fermi surfaces \cite{iida_high-resolution_2004, okubo_unique_2003}. Electronic band structure calculations have predicted strong hybridization of Rh \textit{d} and B \textit{p} bands along the \textit{c}-axis thereby leading to the quasi-one-dimensional nature of the Fermi surfaces\cite{harima_band_2004}. A major contribution to the total density of states is obtained from Rh $4d$ orbitals \cite{chaudhary_role_2023}. Also, strong sensitivity of the superconducting transition temperature on the unit cell size was observed possibly due to the presence of narrow bands at $E_F$. It has been found that a change in lattice constant by 0.001$\AA$ results in a reduction in the superconducting transition temperature by $\sim$ 500mK\cite{chaudhary_role_2023}. 
To understand the low $\Delta$ measured in our experiments that led to an unphysically low value of $\Delta/k_BT_c$, it is important to first understand that PCAR effectively measures a gap amplitude through the measurement of the transport current at the point contacts. For a non-spherical Fermi surface (i.e. anisotropic in $k$-space), the current injected along a direction $\hat{n}$ is given by:
\begin{equation*}
    I \propto \oint_{FS} N_{\vec{k}}(\vec{v_k}.\hat{n}) \,dS_{F} = \langle N_{\vec{k}}v_{k(n)} \rangle_{FS}
\end{equation*}
where $N_{\vec{k}}$ is the density of states, $v_{kn}$ is the component of the Fermi velocity along the direction of current and dS$_{F}$ is an elementary area on the Fermi surface. Therefore, the current through a ballistic point contact is proportional to the projection area of the Fermi surface in a direction perpendicular to that of the current flow. For a superconducting gap $\Delta_{\vec{k}}$, the measured $\Delta$ is such an average parameter given by:
\begin{equation*}
    \langle \Delta \rangle = \frac{\langle \Delta_{\vec{k}}N_{\vec{k}}\vec{v}_{k\hat{n}} \rangle_{FS}}{\langle N_{\vec{k}}\vec{v}_{k\hat{n}} \rangle_{FS}}
\end{equation*}
Hence, if $\Delta_{\vec{k}}$ is anisotropic in the momentum space and is very small along certain momentum directions, then PCAR will measure an average  $\Delta$ that is significantly smaller than the maximum $\Delta$ for the system\cite{aslam_anisotropic_2017}. In this context it is also important to note that previous phonon calculations using density functional perturbation theory have revealed anisotropic phonon density of states\cite{chaudhary_role_2023} where the dominant contributing phonon modes belong to the Kagome network of $Rh$ atoms while the low and high frequency modes originate from the dynamics of the $B$ and $La$ atoms respectively. The idea of an anisotropic order parameter in LaRh$_3$B$_2$ is further validated by the observation of an unreasonably high value of $\Gamma$, the effective broadening parameter. $\Gamma$ accounts not only for the broadening of PCAR spectrum due to quasiparticle lifetime but also for the variation of gap function sensed by the injected current, i.e.
\begin{equation*}
    \frac{\Gamma}{\Delta} \propto \sqrt{\sum_{i}(\Delta_i-\langle \Delta \rangle)^2}
\end{equation*}
Therefore, a large distribution of the gap amplitude in the momentum space leads to a large $\Gamma$ in PCAR experiments\cite{raychaudhuri_evidence_2004}. To note, a visual inspection of the $H_c$-$T_c$ phase diagram reveals an upward curvature \cite{chaudhary_role_2023}. This behavior is usually seen in anisotropic single gap or in multigap systems, and in some unconventional superconductors\cite{zavaritsky_universal_2002, shi_out--plane_2003}. The claim of anisotropic distribution of gap amplitude is also consistent with the specific heat measurements which suggested that $\sim$ 14\% of electrons at the Fermi surface of LaRh$_3$B$_2$ don't participate in superconducting phase transition\cite{chaudhary_role_2023} indicating the presence of low-energy excitations at least along certain momentum directions.  

\section{Conclusion}
In conclusion, we have performed point contact Andreev reflection spectroscopy experiments on the polycrystals of the Kagome superconductor LaRh$_3$B$_2$ and found (a) anomalous suppression of Andreev reflection at the interface, (b) a significant deviation of the experimentally obtained $\Delta$ vs. $T$ from the BCS theory, (c) an abnormally low $\Delta/k_BT_c$, (d) the emergence of an unusually high $\Gamma$. The polycrystalline sample provided us access to the different crystal facets on a single sample surface where the metallic tip was engaged. The distribution of the superconducting energy gap at different points can be understood as a variation of the gap amplitude from one facet to another. All these observations collectively give a strong indication that LaRh$_3$B$_2$ is an anisotropic superconductor.

%\subsection*{Data availability}
%The data that support the findings of this study are available within the article.
%\subsection*{Competing interests}
%The authors declare no competing financial interests.

\section{Acknowledgements}
MG thanks the Council of Scientific and Industrial Research (CSIR), Government of India, for financial support through a research fellowship (Award No. \textbf{09/947(0227)/2019-EMR-I}). NSM and GM thanks UGC for Senior Research Fellowship (SRF). GS acknowledges financial assistance from the Science and Engineering Research Board (SERB), Govt. of India (grant number: \textbf{CRG/2021/006395}).

%% file: mysecondfile.tex
\title{Supporting Information \\
\Large Spectroscopic signature of anisotropic order parameter in Kagome lattice superconductor LaRh$_3$B$_2$}

\author{Mona Garg}
\email{mona@iisermohali.ac.in}
\author{Nikhlesh S. Mehta}
\author{ Ghulam Mohmad}
\author{Savita Chaudhary}
\author{Yogesh Singh}
\author{Goutam Sheet}

\email{goutam@iisermohali.ac.in}
\affiliation{Department of Physical Sciences, Indian Institute of Science Education and Research Mohali, S.A.S. Nagar, Punjab, India}

\maketitle
\section*{Estimating the diameter of point contacts}
Different regimes of transport can be realized at a point contact depending on contact resistance and Knudsen number, $K$ = $l/a$, Where $a$ is the contact diameter, $l$ is the electronic mean free path. To bridge the gap between transport through the ballistic regime (Sharvin resistance) and thermal regime (Maxwell resistance), Wexler provided an interpolation equation\cite{wexler_size_1966}:\par
 \begin{equation}
 R_{PC} = \frac{2h/e^2}{(ak_F)^2} + \Gamma(l/a)\frac{\rho(T)}{2a}    
 \end{equation}
where $\Gamma(l/a)$ is a factor close to unity.\par
where the first one represents the Sharvin resistance and the second one represents the Maxwell resistance.

\newpage
\section*{Addition PCAR spectra}
\begin{figure}[h!]
    \centering
    \includegraphics[width=1\textwidth]{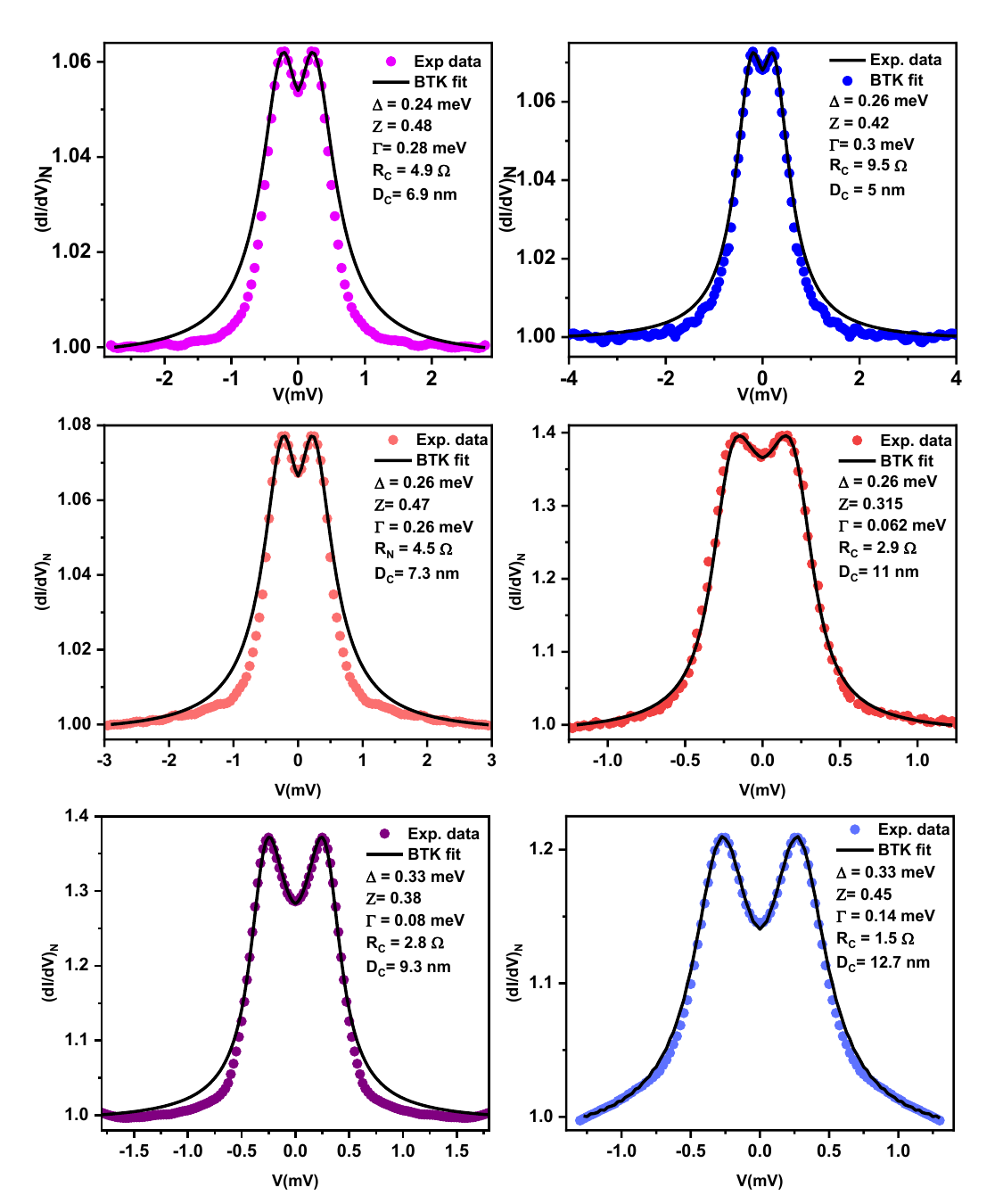}
    \caption{PCAR spectra recorded at different points on the surface of LaRh$_3$B$_2$ using Ag tip and corresponding modified BTK fit (black line).}
\end{figure}
\begin{figure}[h!]
    \centering
    \includegraphics[width=1\textwidth]{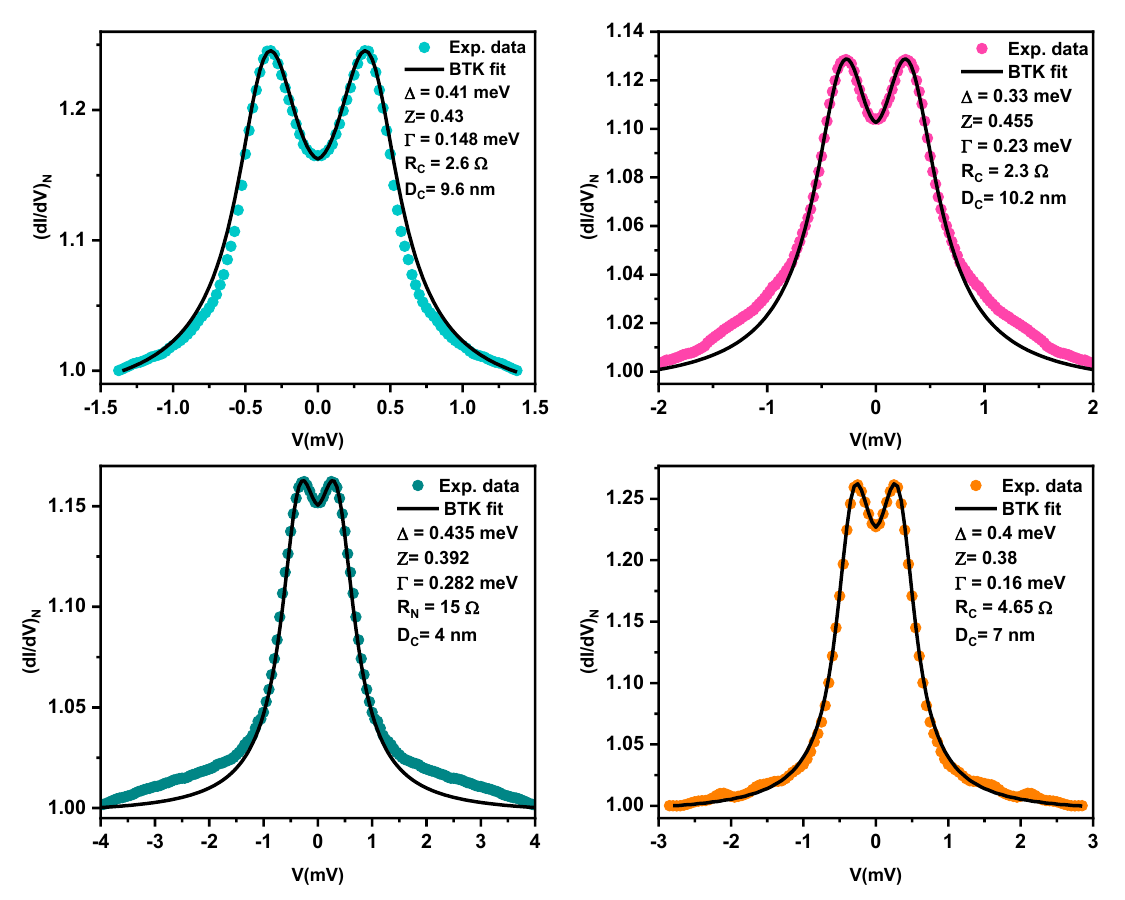}
    \caption{PCAR spectra recorded at different points on the surface of LaRh$_3$B$_2$ using Ag tip and corresponding modified BTK fit (black line).}
\end{figure}
\pagebreak

%% file: main.bbl
\begin{thebibliography}{43}
\bibitem{liu_magnetic_2019} 
D. F. Liu, A. J. Liang, E. K. Liu, Q. N. Xu, Y. W. Li, C. Chen, D. Pei, W. J. Shi, S. K. Mo, P. Dudin, T. Kim, C. Cacho, G. Li, Y. Sun, L. X. Yang, Z. K. Liu, S. S. P. Parkin, C. Felser and Y. L. Chen, Magnetic Weyl semimetal phase in a {Kagomé} crystal, \textit{Science} \textbf{365}, 1282 (2019).

\bibitem{kang_dirac_2020} 
M. Kang, L. Ye, S. Fang, J.-S. You, A. Levitan, M. Han, J. I. Facio, C. Jozwiak, A. Bostwick, E. Rotenberg, M. K. Chan, R. D. McDonald, D. Graf, K. Kaznatcheev, E. Vescovo, D. C. Bell, E. Kaxiras, J. Van Den Brink, M. Richter, M. Prasad Ghimire, J. G. Checkelsky and R. Comin, Dirac fermions and flat bands in the ideal kagome metal {FeSn}, \textit{Nature Materials} \textbf{19}, 163 (2020).

\bibitem{liu_orbital-selective_2020} 
Z. Liu, M. Li, Q. Wang, G. Wang, C. Wen, K. Jiang, X. Lu, S. Yan, Y. Huang, D. Shen, J.-X. Yin, Z. Wang, Z. Yin, H. Lei and S. Wang, Orbital-selective {Dirac} fermions and extremely flat bands in frustrated kagome-lattice metal {CoSn}, \textit{Nature Communications} \textbf{11}, 4002 (2020).

\bibitem{cable_magnetic_1993} 
J. W. Cable, N. Wakabayashi and P. Radhakrishna, Magnetic excitations in the triangular antiferromagnets {Mn$_3$Sn} and {Mn$_3$Ge}, \textit{Physical Review B} \textbf{48}, 6159 (1993).

\bibitem{nakatsuji_large_2015} 
S. Nakatsuji, N. Kiyohara and  T. Higo, Large anomalous {Hall} effect in a non-collinear antiferromagnet at room temperature, \textit{Nature} \textbf{527}, 212 (2015).

\bibitem{hou_observation_2018} 
Z. Hou, W. Ren, B. Ding, G. Xu, Y. Wang, B. Yang, Q. Zhang, Y. Zhang, E. Liu, F. Xu, W. Wang, G. Wu, X. Zhang, B. Shen and Z. Zhang, Observation of {Various} and {Spontaneous} {Magnetic} {Skyrmionic} {Bubbles} at {Room} {Temperature} in a {Frustrated} {Kagome} {Magnet} with {Uniaxial} {Magnetic} {Anisotropy}, \textit{Advanced Materials} \textbf{30}, 1706306 (2018).

\bibitem{ma_rare_2021} 
W. Ma, X. Xu, J.-X. Yin, H. Yang, H. Zhou, Z.-J. Cheng, Y. Huang, Z. Qu, F. Wang, M. Z. Hasan, S. Jia, Rare {Earth} {Engineering} in {RMn$_6$Sn$_6$} ( {R} = {Gd}-{Tm}, {Lu}) {Topological} {Kagome} {Magnets}, \textit{Physical Review Letters} \textbf{126}, 246602 (2021).

\bibitem{pasco_tunable_2019}
C. M. Pasco, I. El Baggari, E. Bianco, L. F. Kourkoutis and T. M. McQueen, Tunable {Magnetic} {Transition} to a {Singlet} {Ground} {State} in a {2D} van der {Waals} {Layered} {Trimerized} {Kagomé} {Magnet}, \textit{ACS Nano} \textbf{13}, 9457 (2019).

\bibitem{zheng_quantum_2024} 
G. Zheng, Y. Zhu, S. Mozaffari, N. Mao, K.-W. Chen, K. Jenkins, D. Zhang, A. Chan, H. W. S. Arachchige, R. P. Madhogaria, M. Cothrine, W. R. Meier, Y. Zhang, D. Mandrus, L. Li, Quantum oscillations evidence for topological bands in kagome metal {ScV}$_{6}${Sn}$_{6}$, \textit{Journal of Physics: Condensed Matter} \textbf{36}, 215501 (2024).

\bibitem{zhang_electronic_2022} 
X. Zhang, Z. Liu, Q. Cui, Q. Guo, N. Wang, L. Shi, H. Zhang, W. Wang, X. Dong, J. Sun, Z. Dun and J. Cheng, Electronic and magnetic properties of intermetallic kagome magnets {R}{V}$_6${Sn}$_6$ ({R} = {Tb}-{Tm}), \textit{Physical Review Materials} \textbf{6}, 105001 (2022).

\bibitem{guguchia_hidden_2023} 
Z. Guguchia, D. J. Gawryluk, S. Shin, Z. Hao, C. Mielke III, D. Das, I. Plokhikh, L. Liborio, J. K. Shenton, Y. Hu, V. Sazgari, M. Medarde, H. Deng, Y. Cai, C. Chen, Y. Jiang, A. Amato, M. Shi, M. Z. Hasan, J.-X. Yin, R. Khasanov, E. Pomjakushina, H. Luetkens, \textit{Nature Communications} \textbf{14}, 7796 (2023).

\bibitem{guin_zerofield_2019} 
S. N. Guin, P. Vir, Y. Zhang, N. Kumar, S. J. Watzman, C. Fu, E. Liu, K. Manna, W. Schnelle, J. Gooth, C. Shekhar, Y. Sun and C. Felser, Hidden magnetism uncovered in a charge ordered bilayer kagome material {ScV$_6$Sn$_6$}, \textit{Advanced Materials} \textbf{31}, 1806622 (2019).

\bibitem{zhang_topological_2017} 
X. Zhang, L. Jin, X. Dai and G. Liu, Topological {Type}-{II} {Nodal} {Line} {Semimetal} and {Dirac} {Semimetal} {State} in {Stable} {Kagome} {Compound} {Mg} $_{3}${Bi}$_{2}$, \textit{The Journal of Physical Chemistry Letters} \textbf{8}, 4814 (2017).

\bibitem{liu_giant_2018} 
E. Liu, Y. Sun, N. Kumar, L. Muechler, A. Sun, L. Jiao, S.-Y. Yang, D. Liu, A. Liang, Q. Xu, J. Kroder, V. Süß, H. Borrmann, C. Shekhar, Z. Wang, C. Xi, W. Wang, W. Schnelle, S. Wirth, Y. Chen, S. T. B. Goennenwein and C. Felser, Magnetic {Weyl} semimetal phase in a {Kagomé} crystal, \textit{Nature Physics} \textbf{14}, 1125 (2018).

\bibitem{lee_quantum-spin-liquid_2007} 
S.-H. Lee, H. Kikuchi, Y. Qiu, B. Lake, Q. Huang, K. Habicht and K. Kiefer, Quantum-spin-liquid states in the two-dimensional kagome antiferromagnets Zn$_x$Cu$_{4-x}$({OD})$_6$Cl$_2$, \textit{Nature Materials} \textbf{6},853 (2007).

\bibitem{zhao_cascade_2021} 
H. Zhao, H. Li, B. R. Ortiz, S. M. L. Teicher, T. Park, M. Ye, Z. Wang, L. Balents, S. D. Wilson, I. Zeljkovic, Cascade of correlated electron states in the kagome superconductor {CsV$_3$Sb$_5$}, \textit{Nature} \textbf{599}, 216 (2021).

\bibitem{chen_roton_2021}
H. Chen, H. Yang, B. Hu, Z. Zhao, J. Yuan, Y. Xing, G. Qian, Z. Huang, G. Li, Y. Ye, S. Ma, S. Ni, H. Zhang, Q. Yin, C. Gong, Z. Tu, H. Lei, H. Tan, S. Zhou, C. Shen, X. Dong, B. Yan, Z. Wang and H.-J. Gao, Roton pair density wave in a strong-coupling kagome superconductor, \textit{Nature} \textbf{599}, 222 (2021).

\bibitem{ortiz_cs_2020} 
B. R. Ortiz, S. M. Teicher, Y. Hu, J. L. Zuo, P. M. Sarte, E. C. Schueller, A. M. Abeykoon, M. J. Krogstad, S. Rosenkranz, R. Osborn, R. Seshadri, L. Balents, J. He and S. D. Wilson, {CsV$_3$Sb$_5$}: {A} $Z_2$ {Topological} {Kagome} {Metal} with a {Superconducting} {Ground} {State}, \textit{Physical Review Letters} \textbf{125}, 247002 (2020).

\bibitem{yu_concurrence_2021} 
F. H. Yu, T. Wu, Z. Y. Wang, B. Lei, W. Z. Zhuo, J. J. Ying and X. H. Chen, Concurrence of anomalous {Hall} effect and charge density wave in a superconducting topological kagome metal, \textit{Physical Review B} \textbf{104}, L041103 (2021).

\bibitem{wen_emergent_2023} 
X. Wen, F. Yu, Z. Gui, Y. Zhang, X. Hou, L. Shan, T. Wu, Z. Xiang, Z. Wang, J. Ying and X. Chen, Emergent superconducting fluctuations in compressed kagome superconductor {CsV$_3$Sb$_5$}, \textit{Science Bulletin} \textbf{68}, 259 (2023).

\bibitem{wu_nature_2021} 
X. Wu, T. Schwemmer, T. Müller, A. Consiglio, G. Sangiovanni, D. Di Sante, Y. Iqbal, W. Hanke, A. P. Schnyder, M. M. Denner, M. H. Fischer, T. Neupert and R. Thomale, Nature of {Unconventional} {Pairing} in the {Kagome} {Superconductors} {A}{V}$_3${Sb}$_5$({A} = {K}, {Rb}, {Cs}), \textit{Physical Review Letters} \textbf{127}, 177001 (2021).

\bibitem{kang_twofold_2022} 
M. Kang, S. Fang, J.-K. Kim, B. R. Ortiz, S. H. Ryu, J. Kim, J. Yoo, G. Sangiovanni, D. Di Sante, B.-G. Park, C. Jozwiak, A. Bostwick, E. Rotenberg, E. Kaxiras, S. D. Wilson, J.-H. Park and R. Comin, Twofold van {Hove} singularity and origin of charge order in topological kagome superconductor {CsV$_3$Sb$_5$}, \textit{Nature Physics} \textbf{18}, 301 (2022).

\bibitem{okubo_unique_2003} 
T. Okubo, M. Yamada, A. Thamizhavel, S. Kirita, Y. Inada, R. Settai, H. Harima, K. Takegahara, A. Galatanu, E. Yamamoto and Y. Nuki, Unique {Fermi} surfaces with quasi-one-dimensional character in {CeRh$_3$B$_2$} and {LaRh$_3$B$_2$}, \textit{Journal of Physics: Condensed Matter} \textbf{15}, L721 (2003).

\bibitem{gui_lair_2022} 
X. Gui and R. J. Cava, {LaIr$_3$Ga$_2$}: {A} {Superconductor} {Based} on a {Kagome} {Lattice} of {Ir}, \textit{Chemistry of Materials} \textbf{34},2824 (2022).

\bibitem{gong_superconductivity_2022} 
C. Gong, S. Tian, Z. Tu, Q. Yin, Y. Fu, R. Luo and H. Lei, Superconductivity in {Kagome} {Metal} {YRu$_3$Si$_2$} with {Strong} {Electron} {Correlations}, \textit{Chinese Physics Letters} \textbf{39}, 087401 (2022).

\bibitem{wang_quantum_2023} 
Y. Wang, H. Wu, G. T. McCandless, J. Y. Chan and M. N. Ali, Quantum states and intertwining phases in kagome materials, \textit{Nature Reviews Physics} \textbf{5}, 635 (2023).

\bibitem{rauchschwalbe_superconductivity_1984} 
U. Rauchschwalbe, W. Lieke, F. Steglich, C. Godart, L. C. Gupta and R. D. Parks, Superconductivity in a {Mixed}-valent {System}:{CeRu$_3$Si$_2$}, \textit{Physical Review B} \textbf{30}, 444 (1984).

\bibitem{mielke_nodeless_2021} 
C. Mielke, Y. Qin, J.-X. Yin, H. Nakamura, D. Das, K. Guo, R. Khasanov, J. Chang, Z. Q. Wang, S. Jia, S. Nakatsuji, A. Amato, H. Luetkens, G. Xu, M. Z. Hasan and Z. Guguchia, Nodeless kagome superconductivity in {LaRu$_3$Si$_2$}, \textit{Physical Review Materials} \textbf{5}, 034803 (2021).

\bibitem{li_anomalous_2011} 
S. Li, B. Zeng, X. Wan, J. Tao, F. Han, H. Yang, Z. Wang and H.-H. Wen, Anomalous properties in the normal and superconducting states of {LaRu$_3$Si$_2$}, \textit{Physical Review B} \textbf{84}, 214527 (2011).

\bibitem{chaudhary_role_2023} 
S. Chaudhary, Shama, J. Singh, A. Consiglio, D. Di Sante, R. Thomale and Y. Singh, Role of electronic correlations in the kagome-lattice superconductor {LaRh$_3$B$_2$}, \textit{Physical Review B} \textbf{107}, 085103 (2023).

\bibitem{10.1063/1.5119372} 
S. Das and G. Sheet, A modular point contact spectroscopy probe for sub-Kelvin applications, \textit{Review of Scientific Instruments} \textbf{90}, 103903 (2019).

\bibitem{jansen_point-contact_1980} 
A. G. M. Jansen, A. P. V. Gelder and P. Wyder, Point-contact spectroscopy in metals, \textit{Journal of Physics C: Solid State Physics} \textbf{13}, 6073 (1980).

\bibitem{PhysRevB.25.4515} 
G. E. Blonder, M. Tinkham and T. M. Klapwijk, Transition from metallic to tunneling regimes in superconducting microconstrictions: Excess current, charge imbalance, and supercurrent conversion, \textit{Phys. Rev. B} \textbf{25} 4515 (1982).

\bibitem{10.1063/1.3479927} 
J. Wei, G. Sheet and V. Chandrasekhar, Possible microscopic origin of large broadening parameter in point Andreev reflection spectroscopy, \textit{Applied Physics Letters} \textbf{97}, 062507 (2010).

\bibitem{PhysRevB.49.10016} 
A. Pleceník, M. Grajcar, Š. Beňačka, P. Seidel, and A. Pfuch Finite-quasiparticle-lifetime effects in the differential conductance of ${\mathrm{Bi}}_{2}$${\mathrm{Sr}}_{2}$${\mathrm{CaCu}}_{2}$${\mathrm{O}}_{\mathit{y}}$/Au junctions, \textit{Phys. Rev. B} \textbf{49} 10016 (1994).

\bibitem{PhysRevB.69.134507} 
G. Sheet, S. Mukhopadhyay and P. Raychaudhuri, Role of critical current on the point-contact Andreev reflection spectra between a normal metal and a superconductor, \textit{Phys. Rev. B} \textbf{69} 134507 (2004).

\bibitem{kumar_nonballistic_2021} 
R. Kumar and G. Sheet, Nonballistic transport characteristics of superconducting point contacts, \textit{Physical Review B} \textbf{104}, 094525 (2021).

\bibitem{iida_high-resolution_2004} 
Y. Iida, S. Souma, T. Sato, T. Takahashi, M. Yamada, T. Okubo, T. Shiromoto, Y. Inada, R. Settai and Y. Onuki, High-resolution angle-resolved photoemission study of {LaRh$_3$B$_2$}, \textit{Physica B: Condensed Matter} \textbf{351}, 271 (2004).

\bibitem{harima_band_2004} 
H. Harima and K. Takegahara, Band structure calculation and Fermi surfaces for {LaRh$_3$B$_2$}, \textit{Journal of Magnetism and Magnetic Materials} \textbf{272-276} 475 (2004).

\bibitem{aslam_anisotropic_2017} 
M. Aslam, S. Gayen, A. Singh, M. Tanaka, T. Yamaki, Y. Takano and G. Sheet, Anisotropic superconductivity in {La}({O},{F}){BiSeS} crystals revealed by field-angle dependent {Andreev} reflection spectroscopy, \textit{Solid State Communications} \textbf{264} (2017).

\bibitem{raychaudhuri_evidence_2004}
P. Raychaudhuri, D. Jaiswal-Nagar, G. Sheet, S. Ramakrishnan, H. Takeya, \textit{Physical Review Letters} \textbf{93}, 156802 (2004).

\bibitem{zavaritsky_universal_2002} 
V. N. Zavaritsky, V. V. Kabanov and A. S. Alexandrov, Universal upper critical field of unconventional superconductors, \textit{Europhysics Letters (EPL)} \textbf{60}, 127 (2002).

\bibitem{shi_out--plane_2003} 
Z. X. Shi, M. Tokunaga, T. Tamegai, Y. Takano, K. Togano, H. Kito and H. Ihara, Out-of-plane and in-plane anisotropy of upper critical field in {MgB}$_2$, \textit{Physical Review B} \textbf{68}, 104513 (2003).

\end{thebibliography}

\begin{thebibliography}{1}
\bibitem{wexler_size_1966}
G. Wexler, The size effect and the non-local Boltzmann transport equation in orifice and disk geometry, \textit{Proceedings of the Physical Society} \textbf{89}, 927 (1966).
\end{thebibliography}
